\pdfoutput=1

\documentclass[11pt]{article}

\usepackage[preprint]{acl}

\usepackage{times}
\usepackage{latexsym}
\usepackage{listings}
\usepackage[T1]{fontenc}

\usepackage[utf8]{inputenc}

\usepackage{rotating}

\usepackage{microtype}
\usepackage{multirow}

\usepackage{inconsolata}
\usepackage{ stmaryrd }

\usepackage{graphicx}

\newcommand{\ppm}[2]{$#1_{\pm{#2}}$}
\newcommand{\tr}[1]{\textbf{#1}}
\newcommand{\emb}[1]{\textbf{#1}}
\usepackage{spverbatim}

\usepackage{amsmath}

\usepackage{todonotes}

\newcommand{\hide}[1]{}

\title{Improving Document Retrieval Coherence for \\Semantically Equivalent Queries}

\author{Stefano Campese \\
  Amazon AGI \\
  University of Trento \\
  \texttt{campeses@amazon.com} \\
  \And
  Alessandro Moschitti \\
  Amazon AGI \\
  \texttt{amosch@amazon.com} \\\And
  Ivano Lauriola \\
  Amazon AGI \\
  \texttt{lauivano@amazon.com} \\}

\begin{document}
\maketitle
\begin{abstract}
Dense Retrieval (DR) models have proven to be effective for Document Retrieval and Information Grounding tasks. 
Usually, these models are trained and optimized for improving the relevance of top-ranked documents for a given query. 
Previous work has shown that popular DR models are sensitive to the query and document lexicon: small variations of it may lead to a significant difference in the set of retrieved documents. 
In this paper, we propose a variation of the Multi-Negative Ranking loss for training DR that improves the coherence of models in retrieving the same documents with respect to  semantically similar queries. 
The loss penalizes discrepancies between  the top-k ranked documents retrieved for diverse but semantic equivalent queries. 
We conducted extensive experiments on various datasets, MS-MARCO, Natural Questions, BEIR, and TREC DL 19/20. The results show that (i) models optimizes by our loss are subject to lower sensitivity, and, (ii) interestingly, higher accuracy.
\end{abstract}

\section{Introduction}

In the recent years, pre-trained Language Models (PLMs) have sown striking performance on a plethora of NLP tasks including, but not limited to, Question Answering, Information Retrieval, Machine Translation, chat-bots, and many more~\cite{min2023recent,wang2023pre}. 
One popular and well-studied application of PLMs is Dense Retrieval (DR)~\cite{karpukhin-etal-2020-dense}, consisting of dual encoders that create dense vector representations (embeddings) of both queries and documents. Embeddings similarities are then used to retrieve relevant documents from an index.

Recently, DR proven to be an effective solution for both, simple document retrieval applications and Retrieval Augmented Generation (RAG)~\cite{zhao2024dense}, where a Larger Language Model (LLM) is tasked to produce answers based on retrieved documents. 
DR models are typically fine-tuned from PLMs to align the embeddings between queries and relevant texts or documents. 
Previous work has shown improvements through various approaches, e.g.: (i) specialized loss functions~\cite{Henderson2017EfficientNL}, (ii) mechanisms to mine meaningful training examples and hard-negatives~\cite{lin-etal-2023-train}, and (iii) labeled data at scale~\cite{nguyen2016ms}. 
%
%
One potential drawback of DR is their \emph{sensitivity} to the query and document lexicon. Intuitively, this is defined as the \emph{difference} in the output response with respect to changing of the query wording~\cite{chen2024unsupervised,liu2023robustness}.
We note two aspects:
First, low query sensitivity is empirically proven to be proportional to high accuracy~\cite{lu-etal-2024-prompts,lauriola-etal-2025-analyzing}. Not being able to answer some variations of the same query corresponds to poor generalization. For instance, a model trained on natural questions may have problems in answering web-like versions of the same queries, or questions with negations~\cite{guo-etal-2025-learning}. 
Second, behavioral studies showed that users start multiple searches with rewritten queries when the initial search output does not contain satisfactory results~\cite{bernard2007defining,jansen2005temporal}, and up to 50\% traffic in early retrieval engines may be just reformulations. Recent work suggests that the problem is not solved in modern retrieval engines~\cite{Wang2021}. 
These aspects may lead to an increase of search cost, as multiple searches require the re-execution of the retrieval pipeline.

%
Previous work explored various approaches to make the model less sensitive, and thus more \emph{coherent}, including synthetic data generation~\cite{guo-etal-2025-learning,chaudhary-etal-2024-relative,meng2022augtriever} and query reformulation~\cite{ma-etal-2023-query}.
The former shows that generating lexical variations of annotated queries can improve the generalization of the model. The latter tries to reshape the query to be more aligned to the DR input while preserving the intent. 
Although query reformulation showed some benefits, it requires the introduction of a \emph{re-writer}~\cite{ma-etal-2023-query}, typically implemented as an LLM, with a consequent drop in efficiency and increase in cost. 

In this work, we focus on analyzing and improving the coherence of DR models, intuitively defined as the ability of a model in retrieving the same set of documents (or the same ranked list) from a given collection (or index) for different lexical variations of the same equivalent input query. 
Differently from most of previous work, based on query reformulation or simple data augmentation, we inject the coherence into the loss function directly. Specifically, we extend the Multiple Negative Ranking (MNR) loss~\cite{Henderson2017EfficientNL} to (i) penalize dissimilarities of embeddings from lexical variations of the same query and to (ii) optimize for query-document similarity alignment. 

To validate the effectiveness of the loss function, we conducted extensive experiments on MS-MARCO, Natural Questions, BEIR, and TREC-DL with multiple PLMs, namely MPNet~\cite{song2020mpnet}, ModernBERT~\cite{warner2024smarter}, and MiniLM~\cite{wang2020minilm}. 
Our results show that our loss consistently improves the coherence of DR models (and thus reducing the general idea of sensitivity to the input query) measured through Rank Biased Overlap (RBO)~\cite{webber2010similarity} between documents retrieved from multiple equivalent queries, with an average increase of +15\% absolute on MS-MARCO, from 0.43 to 0.58, and +29\% on Natural Questions, from 0.38 to 0.67. 
Beyond coherence, our approach shows an improvement in NDCG of +0.60\% MS-MARCO, +1.8\% on NQ, +0.5\% on 11 BEIR, and +1.4\%/0.3\% on TREC-DL benchmarks averaged.


\section{Related work}
\label{sec:related}


\paragraph{Coherence in LLMs}
Popular LLMs have shown to be very sensitive to the input~\cite{voronov-etal-2024-mind,mizrahi-etal-2024-state,arora2022ask,chatterjee-etal-2024-posix},
and the selection of the prompt format plays a crucial role. 
\citet{lu-etal-2024-prompts} demonstrated that coherence can be seen as the opposite of sensitivity, and can be considered as an unsupervised proxy for model performance.
In addition, \citet{raina-etal-2024-llm} performed a deep analysis on adversarial robustness of LLMs, showing how to deceive an LLM judge to manipulate the output and predict inflated scores. 
Except for the input lexicon, the position of words and concepts, e.g.: order of options in multi-choice Q\&A~\cite{zheng2023large}
or order of in-context examples~\cite{liu-etal-2022-makes,zhao2021calibrate},
also affects the judgment. 

Beyond analyzing the phenomenon, \citet{chatterjee-etal-2024-posix} introduced a metric, named POSIX, to measure the prompt sensitivity. 
Moreover, \citet{rabinovich2023predicting} introduced PopQA-TP, a curated dataset that extends PopQA~\cite{mallen2023llm_memorization} with 118,000 paraphrased questions, to benchmark LLMs' sensitivity.
Similarly, \citet{lauriola-etal-2025-analyzing} shower how up to 70B LLMs are unable to provide coherent answers from equivalent queries, and highlighted how coherence optimization is linked to overall accuracy.

\paragraph{Sensitivity in Dense Retrieval}
Narrowing down the focus on Dense Retrieval (DR) models, previous work showed similar insights. 
%
\citet{chen2024unsupervised} proposed an unsupervised technique to make the model scores robust towards irrelevant paragraphs in a document. 
%
\citet{liu2023robustness} studied the sensitivity of models in generative retrieval settings through simple query variations (misspelling, token order modification, rule-based paraphrasing). However, the authors focused on observing the phenomenon and quantify the impact of these simple perturbations. 
Other authors highlighted sensitivity issues from an adversarial viewpoint~\cite{liu2023black,wu2022neural}.


Synthetic query data augmentation has been widely explored as mitigation \cite{chaudhary-etal-2024-relative,liang2020embedding,meng2022augtriever}, showing that generated queries can improve generalization of DR models on some public benchmarks. 
Based on the same intuition, \citet{guo-etal-2025-learning} used query augmentation targeting improvements on queries with negations. 
Similarly,~\citet{sunkara2024enhancing} mixed query data augmentation with multi-task learning. First, they generated variations of queries through back-translation. Then they apply a multi-task loss that forces embeddings of the same queries to be similar while optimizes for query-document relevancy. However, results did not show improvement over classical DR training.

\paragraph{Query re-writing}
As possible mitigation of the coherence issue, query rewriting has become a popular solution, aligning input distribution to DR \emph{favourite} query shape~\cite{he2016learning}. 
For instance, \citet{shi2024enhancing} showed benefits of using multiple re-writing of the query and a subsequent combination of documents retrieved. 
%
On the same line, \citet{ma-etal-2023-query} introduced a trainable rewrite-and-retrieve approach in RAG setting to align the input query to the retriever.
%
%
However, query re-writing requires the introduction of a query generator component in the retrieval pipeline, typically through LLMs, which may cause higher latency and cost in industrial applications. 
Other type of re-writing associated with query expansion~\cite{cao2021automated,baek2025crafting} or conversational Q\&A~\cite{christmann2022conversational, ye-etal-2023-enhancing, qian-dou-2022-explicit,yu2020few} are outside the scope of this work. 
We do not compare against query re-writing approaches as our focus is to train a standalone DR model to improve sensitivity, without external components.



\section{Coherence of ranked documents}

In this section, we introduce our loss that targets sensitivity improvement by penalizing rank inconsistencies with different variations of the query. 

\subsection{Preliminaries - query equivalence}

%

%

%
Let $\mathcal{Q}$ be a distribution of open-domain info-seeking queries and 
let $\mathcal{C}\subseteq\mathcal{Q}$ be a subset of queries equivalent each other, that is, 
$\forall(q_i,q_j)\in\mathcal{C}^2: q_i \equiv q_j$, where $\equiv$ indicates that two questions are semantically equivalent.
In this work, we refer $\mathcal{C}$ as equivalent set or cluster of queries.  
We consider the equivalence definition introduced by \citet{campese-etal-2023-quadro}. 
Two questions $(q_i, q_j)$ are semantically equivalent iff they have the same information-seeking intent and their answers can be interchanged.
In other words, $\forall_a : l(q_i,a) \leftrightarrow l(q_j,a)$, 
where $l$ is a labeling function based on an arbitrary interpretation of correctness. $l(q,a) = 1$ if the answer $a$ is correct for $q$, 0 otherwise.
%
%
Although this definition applies to both, single- and multi-answer queries, this study focuses non-subjective queries with well-defined and verifiable answers. 
When dealing with Q\&A systems or generative LMs, the coherence of the models can be easily defined as the semantic similarity of answers responding to queries belonging to the same cluster~\cite{rabinovich2023predicting}.

In this work, we focus on the coherence of DR models, where their sensitivity is given by the ranked list of relevant documents retrieved. 
Let $\delta$ be a DR scoring model that, given a query $q\in\mathcal{Q}$ and a document $d$ from a given collection $\mathcal{D}$, produces a similarity score, that it $\delta: \mathcal{Q} \times \mathcal{D} \rightarrow [0,1]$. 
For simplicity, we define the top-k list of documents retrieved by $\delta$ from the query $q$ as: 
\begin{equation} \label{eq:rank}
\begin{split}
\psi_{\delta,\mathcal{D}}(q,k) & = [d_{q_1}, d_{q_2},\dots, d_{q_k}] \vspace{0.3cm}\\
s.t. &\text{ }\delta(q, d_{q_i}) \geq \delta(q,d_{q_{i+1}}) \text{ }\forall d_{q_i}\in{\mathcal{D}} 
\end{split}
\end{equation}
Based on this definition of top-k retrieved list of documents, the coherence of a ranking model can easily be defined as the average rank-similarity between multiple queries in a cluster, e.g.:
$\sigma(\psi_{\delta,\mathcal{D}}(q_i,k), \psi_{\delta,\mathcal{D}}(q_j,k))$, where $\sigma$ is a given rank-similarity function and $(q_i,q_j)\in\mathcal{C}^2$ are two queries from the same cluster. 
In this work, we used Rank-Biased Overlap (RBO)~\cite{webber2010similarity} and Spearman correlation, two established metrics to measure similarities of two ranked lists of items.  
The higher the rank-similarity between two equivalent queries, the small the sensitivity of the model to the input. 
Any disparity in the ranks highlights a sensitivity issue.

\subsection{Coherence Ranking Loss}
\label{sec:loss_f}

Here we introduce Coherence Ranking (CR) loss, a support multi-task loss that, paired with classical Multiple-Negative Ranking (MNR) loss, explicitly targets coherence improvements. 

CR loss comprises three main factors: Query Embedding Alignment (QEA), Similarity Margin Consistency (SMC), and query-document relevance implemented through MNR. 
QEA component simply tries to aligning the embeddings of lexically different queries by penalizing their differences measured through Mean Squared Error (MSE).
The second component, SMC, enforces equivalent queries to have the same similarities when compared to the same positive and negative documents. 
Differently from QEA, which focuses on embeddings alignment, SMC targets alignment in similarity scores.
The resulting formulation is:
\begin{align}
 \label{eq:loss}
& \mathcal{L}_{CR} (q, d^+, \mathcal{D}^-, \mathcal{C}) = \nonumber \\
 & \quad \lambda_1 \frac{1}{|\mathcal{C}|} \sum_{q_i\in\mathcal{C}}\|\emb{q}-\emb{q}_i \|^2_2  \nonumber \\
 & + \lambda_2 \sum_{q_i\in\mathcal{C}}\sum_{d\in\mathcal{D}^-} \left( m(q,d^+,d) - m(q_i,d^+,d)\right)^2 \nonumber \\
 & + \text{ } MNR (q, d^+,\mathcal{D}^-),
 \end{align}
where $q\in\mathcal{C}$ is a query from a given cluster, $d^+\in\mathcal{D}$ is a document relevant (or positive) to $q$, 
$\mathcal{D}^-\subset\mathcal{D}$ is a set of irrelevant (or negative) documents and $m(q,d^+,d)$ expresses the difference between the relevance of the two documents (one positive and one negative) with respect to the query, that is:
$
m(q,d^+,d) = s(\emb{q}, \emb{d}^+) - s(\emb{q}, \emb{d}), 
$
where $s$ is a vector similarity function, here implemented as cosine. We use bold symbols to indicate the embeddings associated with queries and documents.

\section{Experiments}\label{exp}
We ran various experiments to evaluate CR loss on MS-MARCO~\cite{nguyen2016ms} and Natural Questions (NQ)~\cite{kwiatkowski2019natural}. Results with BEIR and TREC-DL are reported in Appendix~\ref{apx:beir}.

\paragraph{MS-MARCO.}  This is a popular benchmark for IR. It consists of an index of 8.8M passages documents, 495K training queries, 523K positive query-document pairs, and a number of 5 hard negatives per query extracted as described by \citet{wang2021gpl}. Given that labels of the official test queries are not released, we divided the development set in development and test, with 3490 queries each. For training, we used up to 5 hard-negatives per queries, made available in the official repository. 
\paragraph{Natural Questions (NQ).} It originally contained 132,803 unique queries, each associated with a Wikipedia page used to extract an answer. We were able to successfully extract hard negatives for 120K queries following the technique described by \citet{wang2021gpl}, generating 10 different hard negatives per query. 
    We randomly selected 3,000 queries to create our development set. For the test set we use the original split consisting of 3,452 queries and 2,681,468 passage documents.\\
    
%
For each dataset, MS-MARCO and NQ, we used Phi-3 generate up to 10 different lexical variations of original queries. 
The model is prompted to generate different queries with the same intent and information-seeking need, while varying the writing style. See Appendix~\ref{apx:genq} for the full prompt and some examples of generated queries. The queries are paired with positive documents associated with the input, augmenting the training data. 
A summary of the two datasets is available in Table~\ref{tab:ds}.

\begin{table}[tb]
  \centering
  \begin{tabular}{lcc}
    \hline
    \textbf{} & \textbf{MS-MARCO} & \textbf{NQ}\\
    \hline
    Queries TRAIN & 495260 & 119554 \\
    Queries DEV & 3490 & 3000 \\
    Queries TEST & 3490 & 3452 \\
    Hard negatives	& 5 & 10 \\
    Gen. Queries & 10 & 10 \\
    \hline
  \end{tabular}
  \caption{Datasets statistics.}
  \label{tab:ds}
\end{table}

In most of our experiments, we considered the following baselines and configurations that are derived from MPNet~\cite{song2020mpnet}:
\begin{description}
    \item[Public checkpoint] - As simplest baseline we consider the public checkpoint continuously pre-trained on various supervised and self-supervised Sentence Text Similarity (STS) tasks including, but not limited to, paraphrasing, question answering, information retrieval, and natual language inference\footnote{Public checkpoint available at \url{https://huggingface.co/sentence-transformers/multi-qa-mpnet-base-cos-v1}.}. 
    \item[Fine-Tuning] - The public checkpoint is fine-tuned on target training data, that is MS-MARCO or NQ. Training data consists of triplets $\langle q, d^+, \mathcal{D}^-\rangle$, where $q\in\mathcal{Q}$ is a query, $d^+$ is a relevant document, and $\mathcal{D}^-$ is the set of hard negatives associated with $q$. Following the established training approach of DRs, MNR loss is employed.
    \item[Query Augmentation] - The training data is expanded with the equivalent but lexically different queries generated through Phi. For each training triplet $\langle q, d^+, \mathcal{D}^-\rangle$ we consider 10 extra examples $\{\langle q_i, d^+, \mathcal{D}^-\rangle\}_{i=1}^{10}$, where $q_i$ is an equivalent query generated from $q$.
    \item[$\mathcal{L}_{QQ}$] - Generated queries are used to enforce query similarity reasoning, replacing data augmentation. The training mixes query/document and query/generated batches in round-robin fashion (multi-task learning). 
    On each iteration, we apply (i) an optimization step with simple MNR as described in the FT approach; (ii) a second optimization step where we optimize for query similarity, training examples consist of $\langle q_i, q_j \rangle$, $q_i \equiv q_j$. 
    This baselines shows the impact of training the model to learn similarities over different queries, improving the rank indirectly. 
    \item[$\mathcal{L}_{CR}$] - We used our loss as defined in Section~\ref{sec:loss_f}, Eq.~\ref{eq:loss}, that jointly optimizes over MNR and query-similarity. 
    \item[Full] - We used $\mathcal{L}_{CR}$ and query augmentation. 
    \item[Lexical] - We also considered two additional lexical baselines, BM25 and SPLADE-v3~\cite{lassance2024splade}. For MS-MARCO we used the BM25 corpus from Pyserini~\cite{Lin_etal_SIGIR2021_Pyserini}, which already includes query expansion\footnote{Pyserini MS-MARCO corpus: msmarco-v1-passage.d2q-t5-docvectors}. 
\end{description}

Amongst other publicly available models, we selected MPNet as (i) it has not a prohibitive dimension (100M parameters) that could affect the volume of our experiments and (ii) it showed leading performance compared to models of similar size on various IR benchmarks as reported in the Sentence Transformer framework\footnote{Public leaderboard as January 2025 \url{https://www.sbert.net/docs/sentence_transformer/pretrained_models.html}.}. However, other models are tested in Section~\ref{sec:newmodels} to assess generalization of our approach. 
\paragraph{Training details} For each configuration, we used the validation set to find the best configuration of hyper-parameters, including learning rate $\{5/7\cdot 10^{-6}, 1/2/3 \cdot 10^{-5}\}$, and batch size $\{2^x\}_{x=4}^{10}$. We used AdamW optimizer and warmup rate of 10\% of the total training steps. 
We set a limit of 15 epochs for training with an early stopping and a patience of 5. The loss coefficients $\lambda_1$ and $\lambda_2$, which balance the different components of our objective function, were evaluated across $\{0, 0.2, 0.5, 0.8, 1\}$ to determine their optimal values.
For models training, we utilized 8 NVIDIA H100 GPUs. 

\paragraph{Metrics and evaluation} We consider two sets of metrics to evaluate the relevance of top-k retrieved documents and the coherence of the models when answering lexical variations of equivalent questions. 
To measure document relevance, we used standard IR metrics: P@1, NDCG@10, MRR@10, and MAP@100. To evaluate relevance, we used only test queries from the original split (no generated queries). 
Regarding the coherence of the models, we fist run the models on all generated queries (10 per each input test query). Then, without accounting for labels, we compared the rank produced by the original query and the ones produced by generated queries. To measure the alignment and average rank-similarity between original and generated queries, we used RBO~\cite{webber2010similarity} and Spearman metrics. For simplicity, we considered the top-5 ranked items. The rank similarity is averaged across all test queries. The higher the rank correlation, the higher the coherence of the model, i.e. its ability of generating the same rank while prompted with different input variations. 

\subsection{Main results}

\begin{table*}
  \centering
  \begin{tabular}{lcccccc}
    \hline
    \textbf{Model} & \textbf{P@1} & \textbf{NDCG@10} & \textbf{MRR@10} & \textbf{MAP@100} & \textbf{RBO@5} & \textbf{Spearman@5}\\
    \hline
    \multicolumn{7}{c}{MS-MARCO} \\
    \hline
Public ckpt          &	21.58 &	39.88&	33.79&	34.27 & \ppm{0.42}{0.25} & \ppm{0.46}{0.12}\\
FT	        & \ppm{22.82}{0.11}	& \ppm{41.51}{0.08}	& \ppm{35.34}{0.12}	& \ppm{35.68}{0.11} & \ppm{0.46}{0.26} & \ppm{0.47}{0.13}\\
+ Q. Augm.	& \ppm{21.85}{0.12}	& \ppm{40.05}{0.21}	& \ppm{33.96}{0.41}	& \ppm{34.31}{0.21} & \ppm{0.59}{0.27} & \ppm{0.54}{0.17}\\
+ $\mathcal{L}_{QQ}$	    & \ppm{22.87}{0.21}	& \ppm{41.31}{0.10}	& \ppm{35.10}{0.08}	& \ppm{35.50}{0.10} & \ppm{0.51}{0.27} & \ppm{0.49}{0.15}\\
+ $\mathcal{L}_{CR}$	    & \ppm{\tr{23.01}}{0.10}	& \ppm{\tr{41.98}}{0.17}	& \ppm{\tr{35.73}}{0.16}	& \ppm{\tr{35.70}}{0.13} & \ppm{0.60}{0.26}& \ppm{0.53}{0.17} \\
Full                  & 22.46&	41.43&	 34.71&	  35.18 & \ppm{\tr{0.63}}{0.26}& \ppm{\tr{0.55}}{0.18}\\

\hline
BM25 & 16.74&	33.19&	27.13&	27.85 & \ppm{0.22}{0.24}& \ppm{0.45}{0.11}\\
SPLADE-v3 & 21.74&	40.08&	33.72&	34.35 & \ppm{0.46}{0.28} & \ppm{0.49}{0.15}\\
    \hline
    \multicolumn{7}{c}{Natural Questions} \\
    \hline
    Public ckpt            &	30.71&	46.53&	42.59&	40.79&   \ppm{0.57}{0.22}&    \ppm{0.49}{0.15}\\
FT & \ppm{38.16}{0.17}	& \ppm{52.16}{0.13}	& \ppm{49.50}{0.17} & \ppm{47.50}{0.18} &   \ppm{0.54}{0.23}&	 \ppm{0.49}{0.16}\\
+ Q. Augm.  & \ppm{38.57}{0.11}	& \ppm{53.0}{0.01}	& \ppm{49.89}{0.08}	& \ppm{47.66}{0.16}&   \ppm{0.66}{0.23}&	  \ppm{0.54}{0.19}\\
+ $\mathcal{L}_{QQ}$ & \ppm{38.84}{0.04}	& \ppm{53.42}{0.07}	& \ppm{50.23}{0.08}	& \ppm{48.25}{0.10}&    \ppm{0.59}{0.23}&	    \ppm{0.51}{0.17}\\
 + $\mathcal{L}_{CR}$ & \ppm{\tr{39.49}}{0.11}	& \ppm{\tr{53.85}}{0.08}	& \ppm{\tr{50.65}}{0.09}	& \ppm{\tr{48.56}}{0.04}&   \ppm{0.70}{0.22}&		\ppm{0.55}{0.19}\\
Full                        &	39.36&	53.73&	50.50&	48.29&   \ppm{\tr{0.71}}{0.21}&		\ppm{\tr{0.57}}{0.20}\\

\hline
BM25                        &	16.48&	30.55&	26.34&	25.86&   \ppm{0.40}{0.27}&		\ppm{0.49}{0.15}\\
SPLADE-v3                   &	29.66&	44.89&	41.11&	39.43&   \ppm{0.65}{0.23}&	    \ppm{0.54}{0.18}\\

    \hline
  \end{tabular}
  \caption{Results on MS-MARCO and NQ. Best results are highlighted in bold. RBO and Spearman measure the rank-correlation, and thus the coherence of the models. Results are averaged across 5 different runs.}
  \label{tab:main_results}
\end{table*}

Table~\ref{tab:main_results} reports the performance, in terms of document relevance (P@1, NDCG@10, MRR@10, MAP@100) and coherence (RBO@5, Spearman@5), of all baselines and our proposed approach as described in Section~\ref{exp}. 
The table shows multiple key insights:
First, the adoption of generated queries (through Phi-3 as described in Appendix~\ref{apx:genq}) to teach the model working with different input variations, either in form of data augmentation (see \textbf{Q. Augmentation} in the table) or question similarity loss ($\mathcal{L}_{QQ}$), shows inconsistent results. 
When used in MS-MARCO training, generated queries produced a drop in document relevancy metrics (e.g.: -1.46 and -0.20 NDCG@10 with Query Augmentation and $\mathcal{L}_{QQ}$ respectively). 
However, the same techniques lead to an improvement in NQ (e.g.: +0.84 and +1.26 NDCG@10). We hypothesize that generated queries are a mixed blessing and this behavior is linked to the volume of the training data. 
On the one hand, generated queries can space out the data from the test distribution, leading to lower results in MS-MARCO.  
On the other hand, NQ is smaller and thus additional generated data may have higher importance and contribute to metrics improvement. 
Our proposed approach ($\mathcal{L}_{CR}$) shows better results on both datasets on all relevance metrics. 
It is worth to notice that the combination of data augmentation and coherence loss (\textbf{Full}) does not show benefits in document relevance, suggesting that our mechanism to train on query variations is superior to simple query augmentation.  
%
%
Regarding the coherence, we observed that generated queries lead to a strong and consistent improvement in both, RBO and Spearman correlations, over simple fine-tuning. This is expected as the models are explicitly trained to align equivalent yet different questions to the same embedding space or to enforce similarities between different queries and the same documents. 
Although query augmentation is a surprisingly strong baseline for ranking coherence, our proposed approach showed better results. 

A final highlight goes to lexical baselines (\textbf{BM25}, \textbf{SPLADE-v3}). Not surprisingly, BM25 is the least coherent approach. The technique, entirely based on tokens overlap, produces results that are tailored to the input wording. 
Differently, SPLADE, thanks to its ability of highlighting the most relevant tokens and entities, showed better coherence, comparable to dense retrieval baselines.


Other experiments comparing our approach against reformulation strategies are described in Appendix~\ref{apx:reformulation}.

\subsection{Ablation study on loss components}
As described in Section~\ref{sec:loss_f}, our proposed loss comprises two components. 
The first penalizes embedding misalignment between different variations of the same query, enforcing the embeddings to be query-shape agnostic. 
The second acts on the margins, and enforces equivalent queries to have the same distance with positive and negative documents.
The combination of these two components led to the improvement showed in the previous results. 
Note that our loss does not replace MNR,  it extends it through an additional penalty factor. 
Table~\ref{tab:ablation_loss} shows document relevance and ranking coherence while using individual components of the loss in addition to standard MNR.
Results highlight that the combination of query embeddings alignment and margin consistency is the key aspect, and individual components do not produce the same improvement. 

\begin{table}[tb]
  \centering
  \begin{tabular}{lccc}
    \hline
    \textbf{Loss} & \textbf{P@1} & \textbf{NDCG@10} & \textbf{RBO@5}\\
    \hline
    \multicolumn{4}{c}{MS-MARCO} \\
    \hline
$\mathcal{L}_{QEA}$ & 22.78&	41.26&	\ppm{0.20}{0.16}\\
$\mathcal{L}_{SMC}$ & 22.81&	41.51&	\ppm{0.22}{0.18}\\ 
$\mathcal{L}_{CR}$ 
                      & \tr{23.01}&  \tr{41.57}&   \ppm{\tr{0.34}}{0.24}\\
    \hline
    \multicolumn{4}{c}{Natural Questions} \\
    \hline
$\mathcal{L}_{QEA}$         &	38.12&    51.63&  \ppm{0.66}{0.23}   \\
$\mathcal{L}_{SMC}$         &	38.88&    53.22&  \ppm{0.57}{0.23}\\
$\mathcal{L}_{CR}$ 
                            &	\tr{39.54}&	\tr{53.92}&	  \ppm{\tr{0.70}}{0.22}\\
   \hline
  \end{tabular}
  \caption{Ablation study on loss components: Query Embedding Alignment and Similarity Margin Consistency. RBO measures ranking consistency.}
  \label{tab:ablation_loss}
\end{table}

\subsection{Models generalization study}
\label{sec:newmodels}

\begin{table*}
  \centering
  \begin{tabular}{lcccccc}
    \hline
    \multirow{2}{*}{\textbf{Configuration}} & \multicolumn{3}{c}{\textbf{MiniLM-v2-12L}} & \multicolumn{3}{c}{\textbf{ModernBERT-base}}\\
     & \textbf{P@1} & \textbf{NDCG@10} & \textbf{RBO@5} & \textbf{P@1} & \textbf{NDCG@10} & \textbf{RBO@5}\\
    \hline
    \multicolumn{7}{c}{MS-MARCO} \\
    \hline
Public ckpt          &	    21.6    &39.1   &\ppm{0.39}{0.24} & 15.0 & 31.0 &\ppm{0.40}{0.25}\\
FT                    &	    22.6    &40.5   &\ppm{0.44}{0.26} & 22.8 & 41.6 &\ppm{0.39}{0.25}\\
+ Q. Augm.             &	    22.7    &40.4   &\ppm{0.55}{0.27} & 21.7 & 39.9 &\ppm{0.56}{0.26}\\
+ $\mathcal{L}_{QQ}$  &	    22.8    &40.5   &\ppm{\tr{0.57}}{0.27} & 21.9 & 40.6 &\ppm{0.49}{0.26}\\
+ $\mathcal{L}_{CR}$ 
                      &     \tr{23.3}    &\tr{41.1}   &\ppm{\tr{0.57}}{0.27} & \tr{23.0} & \tr{41.9} &\ppm{\tr{0.56}}{0.26}\\
    \hline
    \multicolumn{7}{c}{Natural Questions} \\
    \hline
    Public ckpt            &	26.3  &41.4   &\ppm{0.53}{0.23} & 21.8 & 37.6 &\ppm{0.58}{0.23}\\
FT                          &	34.8  &48.3   &\ppm{0.46}{0.23} & 36.6 & 50.4 &\ppm{0.15}{0.19}\\
+ Q. Augm.                   &	35.4  &48.1   &\ppm{0.61}{0.25} & 35.9 & 50.2 &\ppm{0.61}{0.23}\\
+ $\mathcal{L}_{QQ}$        &	35.4  &48.7   &\ppm{0.44}{0.24} & 36.8 & 51.0 &\ppm{0.38}{0.24}\\
+ $\mathcal{L}_{CR}$ 
                            &	\tr{36.1}  &\tr{49.2}   &\ppm{\tr{0.65}}{0.22} & \tr{37.2} & \tr{51.1} &\ppm{\tr{0.65}}{0.22}\\
    \hline
  \end{tabular}
  \caption{Document relevance and coherence of MiniLM and ModernBERT. RBO measures the coherence.}
  \label{tab:results_other_models}
\end{table*}

All previous experiments were based on MPNet due to its performance on various IR benchmarks compared to models of similar size (approx 100M parameters). 
To stress the generalization of the proposed loss, we tested the latter on other two popular transformer models: MiniLM-v2-12L and ModernBERT-base. 
MiniLM is a efficient yet effective solution for dense retrieval. It consists of 33M learnable parameters only. We considered the checkpoint pre-trained for STS\footnote{\url{sentence-transformers/all-MiniLM-L12-v2}.}. 
ModernBERT is a recent model designed for long sequences. We considered the base version consisting of 133M parameters\footnote{\url{answerdotai/ModernBERT-base}.}. 
Given that ModernBERT was simply trained with MLM objective, we continuously trained the checkpoint on 1.5B text-similarity pairs, following the same STS training applied to MPNet and MiniLM. 
Details of the training are available in Appendix~\ref{apx:sts}. 

Results for a restricted set of configurations are showed in Table~\ref{tab:results_other_models}. 
Remarkably, both models show the same trend previously observed with MPNet. Our proposed loss improves both, the document relevance and the rank coherence, on both datasets. 
These results suggest how our loss generalizes over multiple models and is not tailored to a specific solution. 
The effect of our STS pretraining in ModernBERT is further analyzed in Appendix~\ref{apx:mbert}.

\subsection{Retrieve and Rank evaluation}
\label{sec:exp_rr}
DR is often a component of a more complex Question Answering or Chat pipeline. Typically, retrieve and re-rank or retrieve and generate solutions are adopted, where the top-k documents selected by a dense retrieval are further re-ranked or used as part of LLM grounding to generate an answer. 
Although the order of documents as input of LLM is important, as discussed in Section~\ref{sec:related}, the same becomes irrelevant in retrieve and re-rank pipelines as document re-rankers typically produce scores to each document that do not depend on the retrieval position. 
%
As long as the retrieval model can retrieve the same set of documents from different lexical variations of the input, then a document re-ranker can potentially select the same content. 

To explore further this aspect, we simulated a retrieve and re-rank application where a document ranking cross-encoder takes the top-50 documents selected by a DR model, re-ranks them, and selects the most relevant one. 
%
Let $\psi_{\delta,\mathcal{D}}(q,k)$ (see Eq.~\ref{eq:rank}) be the set of top-k documents (in our experiment, k=50) retrieved by a given retrieval model from a test query $q$ of a certain cluster $\mathcal{C}$.
Let $d^*\in\psi_{\delta,\mathcal{D}}(q,k)$ be the document selected by the re-ranker.
We define as \emph{re-ranking opportunity} the probability of $d^*$ to appear in the top-k documents retrieved from any other lexical variation of the input query belonging to the same cluster $\mathcal{C}$:
$opportunity(q) = \frac{1}{|\mathcal{C}|} \sum_{q_i\in\mathcal{C}} \textbf{1}_{\psi_{\delta,\mathcal{D}}(q_i,k)}(d^*),$
where $\textbf{1}$ is the indicator function. 
%
Given the best selection from the re-ranker, the re-ranking opportunity measures the likelihood that the same selected document would be made available by the retrieval while prompted with different equivalent questions. In this sense, the reranker has the same opportunity of selecting the same or a better document.
The higher the opportunity the lower the possibility of dropping the highest re-ranked document due to a sensitivity issue. 
Table~\ref{tab:results_rr} shows the re-ranking opportunity on MS-MARCO and NQ while using a state-of-the-art document re-ranker\footnote{\url{https://huggingface.co/BAAI/bge-reranker-large}}.

\begin{table}
  \centering
  \begin{tabular}{lcccc}
    \hline
    \textbf{Configuration} & \rotatebox{90}{\textbf{MPNet}} & \rotatebox{90}{\textbf{MiniLM}} & \rotatebox{90}{\textbf{Mod. BERT}} & \rotatebox{90}{\textbf{M.B. {\footnotesize w/o STS}}}\\
    \hline
    \multicolumn{5}{c}{MS-MARCO} \\
    \hline
Public ckpt          &	    75.7&   73.4   & 74.9   & -      \\
FT                    &	    79.7&   78.4   & 75.8   & 71.0    \\
+ Q. Augm.             &	    85.9&   83.7   & 84.5   & 83.1   \\
+ $\mathcal{L}_{QQ}$  &	    82.4&   80.9   & 83.0   & 80.2   \\
+ $\mathcal{L}_{CR}$ 
                      &     \tr{87.0}&   \tr{85.7}   & \tr{85.5}   & \tr{84.0} \\
BM25                  &     \multicolumn{4}{c}{59.4}       \\
SPLADE-v3             &     \multicolumn{4}{c}{77.7}       \\
    \hline
    \multicolumn{5}{c}{Natural Questions} \\
    \hline
    Public ckpt            &	59.5&   31.9   & 59.3 & -\\
FT                          &	58.9&   52.6   & 11.8 & 7.8\\
+ Q. Augm.                  &	67.5&   63.6   & 59.4 & 50.5\\
+ $\mathcal{L}_{QQ}$        &	55.4&   66.0   & 23.7 & 25.1\\
+ $\mathcal{L}_{CR}$ 
                            &	\tr{70.9}&  \tr{70.4}    & 65.8 & 54.6\\
BM25                        &   \multicolumn{4}{c}{63.2}       \\
SPLADE-v3                   &   \multicolumn{4}{c}{67.5}       \\
    \hline
  \end{tabular}
  \caption{Re-ranking opportunity, how many times the best re-ranked document is retrieved in the top-50 documents from different variations of the query. BGE model was used as re-ranker (cross-encoder).}
  \label{tab:results_rr}
\end{table}

The re-ranking opportunity showed in the table aligns to other results. Our proposed loss makes the model more coherent beyond simple retrieval. All retriever  tested, beyond simple document relevancy, have higher chances to retrieve the best selection from the re-ranker, regardless the shape of the query in input. 
Compared to simple Fine-Tuning, our losses increases the opportunity by 9.3\% on average (8.1\% if we exclude ModernBERT without STS training).
Note that this experiment does not indicate whether a new/different top-ranked document is relevant or not. Here, we just highlight that the new top-1 document that the retrieve and re-rank pipeline would select with different variations of the input is different, but not necessarily worse or better. 

Other findings on a simple retrieve \& generate application are discussed in Appendix~\ref{apx:randg}.

\subsection{Retrieval complexity}
\label{exp:complexity}

We hypothesize that coherent models are particularly valuable for queries where multiple documents share similar relevance scores. To investigate this, we focused our analysis on original queries (non-generated) from MS-MARCO and NQ where the difference in retrieval scores between the top-1 and 50th ranked document was less than 0.1. Such cases represent complex information needs where the retrieval task becomes particularly challenging, as multiple documents exhibit comparable relevance to the query with minimal score differences. For instance, in MS-MARCO, we observed this phenomenon with queries like \textit{"What constitutional amendment granted American women suffrage?"}, \textit{"Can you describe the gallbladder's position in the human anatomy?"} and \textit{"What is the specific location for viewing the total solar eclipse?"}. Similarly, in NQ, queries such as \textit{"where does the great outdoors movie take place"} and \textit{"who was the declaration of independence written for"} demonstrated this characteristic. In these cases, multiple documents received nearly identical relevance scores, making the final ranking highly sensitive to small score variations. This underscores the importance of maintaining coherence in the ranking process, as minimal differences in retrieval scores can significantly impact the final document order.

\begin{table}[tb]
\begin{tabular}{lcc}
\hline
   \textbf{Configuration}                 & \textbf{MS-MARCO}         & \textbf{NQ}              \\

\hline
Public ckpt             & \ppm{0.16}{0.14} & \ppm{0.41}{0.21} \\
FT                & \ppm{0.17}{0.14} & \ppm{0.25}{0.17} \\
+ Gen. Qs            & \ppm{0.32}{0.23} & \ppm{0.43}{0.24}  \\
+ $\mathcal{L}_{QQ}$  & \ppm{0.24}{0.18} & \ppm{0.30}{0.20} \\
+ $\mathcal{L}_{CR}$             & \ppm{0.34}{0.24} & \ppm{0.49}{0.23} \\
Full       & \ppm{0.38}{0.25} & \ppm{0.52}{0.24} \\
BM25                & \ppm{0.07}{0.14} & \ppm{0.36}{0.27} \\
SPLADE\_V3          & \ppm{0.23}{0.21} & \ppm{0.48}{0.26} \\
\hline
\end{tabular}
  \caption{RBO@5 (coherence) on a subset "most complex" queries, i.e. queries where the retrieval score of the top-1 and the 50-th document is differs less than 0.1.}
  \label{tab:complex}
\end{table}

Table~\ref{tab:complex} shows the results of this evaluation. As expected, the coherence measured through RBO is generally much lower compared to the full set (see Table~\ref{tab:main_results}), corroborating our conjecture on the retrieval complexity.
Our proposed loss has a drastic contribution, especially on MS-MARCO, improving coherence from 0.16 to 0.38 (+138\% relative).

\section{Conclusions}
This work analyzes the ranking-coherence of Dense Retrieval (DR) models, that is their ability of retrieving the same content when prompted with different lexical variations of the input query. 
Our experiments show that classical FT based on popular losses and hard-negatives mining leads to poor coherence. As countermeasure, simple data augmentation or multitask training (query-query similarity and query-document alignment) have proven to increase coherence while keeping comparable accuracy. 
On top of that, our loss function, which jointly (i) penalizes embeddings distance between equivalent queries and (ii) enforces margin between different queries and the same positive/negative documents to be the same, further improves both, accuracy and coherence.
Our results, conducted on multiple benchmarks by using different models indicates  high generalization.

\section{Limitations}
The main focus of this work is the coherence of DR models. However, DRs are just a component of state-of-the-art pipelines based on retrieval (typically lexical+dense) and LLMs to generate answers. 
How DR coherence affects the entire pipeline is not deeply explored in this work. Experiments in Section~\ref{sec:exp_rr} show early evidence on how a state-of-the-art document re-ranker may take benefits from a more coherent DR. However, coherence of the re-ranker itself is outside the scope of this work. 
Regarding LLMs' coherence, related work~\cite{lauriola-etal-2025-analyzing} showed that popular models are poorly coherent, and the input query shape heavily affects the final result. 
Based on these premises, an exhaustive evaluation of an end-to-end pipeline requires different work outside the scope of this paper. 

The improvement of document relevancy may seem limited: (i) +0.14 P@1 and +0.47 NDCG@10 on MS-MARCO, (ii) +0.65 P@1 and +0.43 NDCG@10, (iii) +0.48 NDCG@10 on BEIR (average across IR tasks), and +0.68/+0.21 NDCG@10 on TREC-DL (Appendix~\ref{apx:beir}) from the best baseline. 
As discussed before, one main motivation behind coherence optimization is based on previous work evidence, where more coherent models are showed to improve relevancy by \emph{recovering} errors from \emph{unfavorable} input shape. 
However, although we observed a significant improvement in ranking overlap, the same improvement is not directly translated into relevance.
It is worth to notice that the desired outcome from this work is not a accuracy improvement but producing models with higher coherence. As mitigation, we would like to highlight that all experiments conducted on multiple datasets (MS-MARCO, NQ, 11 BEIR benchmarks, and TREC-DL) with different models (MPNet, MiniLM, ModernBERT with and without STS training) are aligned and show similar trends. 


Finally, our results in Section~\ref{exp:complexity} suggest that coherence may gain importance in scenario where there are many similar documents, where small input differences can cause a drastic change in the retrieval score, and thus the final rank. This evidence raises the question, how does coherence impact real-world applications, based on web indexes with Billions of documents?

\bibliography{custom}

\appendix


\section{Queries generation}
\label{apx:genq}
To generate equivalent queries, we uses Phi-3\footnote{\url{microsoft/Phi-3-mini-128k-instruct}}, 3.8B parameters. For each query in the MSMARCO and NQ datasets, we generated 10 semantically equivalent reformulations, maintaining the original intent while introducing linguistic diversity.

The generation process utilized the following prompt:

\small
\begin{spverbatim}
You are a powerful question rephraser and question generation system.
Given a question coming from the MSMARCO dataset, your task is to generate 10 EQUIVALENT questions using different styles coming from other different datasets.
Two questions are defined EQUIVALENT, is they (i) are asking for exact same thing even if they contain a very different wording, and (ii) they require the same answer.

You can use the following dataset styles:
    - SQuAD-style: Starts with an introductory phrase and focuses on a specific piece of information.
    - MS-MARCO-style: Framed as a request for information, with a more open-ended tone.
    - DuoRC-style: Asks about typical or common symptoms/indicators, using "if" to set up the context.
    - HotpotQA-style: Combines a request for key indicators with a follow-up on how to identify them.
    - NQ-style question: Concise and direct, focused on a specific piece of information. Typically starts with question words like "what", "who", "where", etc.
    - TriviaQA-style question: More open-ended and sometimes requires more detailed or nuanced answers. May include additional context or keywords to guide the response.
    - WebQA-style question: Framed as a request for a list or set of information related to the topic. Often starts with phrases like "Can you provide...", "List the...", or "Identify the..."

Your task is to produce a well formatted and parsable JSON containing the EQUIVALENT questions. 
The produced output must be EXACTLY AS FOLLOWS:
```\{
    "original\_question": \$INPUT\_QUESTION,
    "equivalent\_questions":[ 
        \{'question': \$EQUIVALENT\_QUESTION\_1, 'style': \$EQUIVALENT\_QUESTION\_STYLE\_1\}, 
        \{'question': \$EQUIVALENT\_QUESTION\_2, 'style': \$EQUIVALENT\_QUESTION\_STYLE\_2\},
        \dots,
        \{'question': \$EQUIVALENT\_QUESTION\_10, 'style': \$EQUIVALENT\_QUESTION\_STYLE\_10\}
    ],
\}```

Where the \$EQUIVALENT\_QUESTION\_NTH and \$EQUIVALENT\_QUESTION\_STYLE\_NTH are the generated question and the used style.

To produce the JSON you MUST respect the following rules:
+ The generated questions should be short and concise when possible.
+ Remember: two questions are equivalent if (i) they are asking for exact same thing, and (ii) they require the same answer.
+ Remember: each generated question must follow a different style.
+ Remember: the output must be a valid JSON ready to be used without further post-processing.

Here you can find an example:
INPUT\_QUESTION:
symptoms of a dying mouse

OUTPUT JSON:
\{
    "original\_question": "symptoms of a dying mouse",
    "equivalent\_questions":[
        {"question":"What are the typical symptoms that indicate a mouse is dying?", "style":"NQ"}, 
        {"question":"Identify the most common signs that a mouse is approaching the end of its life.", "style":"TriviaQA"},
        {"question":"Can you provide a list of the primary indicators that a mouse is in the process of dying?", "style":"WebQA"},
        {"question":"According to medical experts, what are the primary symptoms that indicate a mouse is nearing the end of its life?", "style":"SQuAD"},
        {"question":"I need to know the most common signs that a mouse is dying. Can you provide me with that information?", "style":"MS-MARCO"},
        {"question":"What are the key indicators that a mouse is in the process of dying, and how can these be identified?", "style":"HotpotQA"},
        {"question":"If a mouse is showing signs of dying, what are the typical symptoms that would be observed?", "style":"DuoRC"},
        {"question":"How does the appearance of a mouse's coat change when it's approaching death?", "style":"NQ"},
        {"question":"What changes in eating and drinking habits suggest a mouse is near death?", "style":"NQ"},
        {"question":"As a mouse approaches death, it may show this sign related to body temperature. What is it?", "style":"TriviaQA"}
    ]
\}

Remember, just return the JSON, no additional text.

Here is the input:
{question}

Please provide your JSON output
\end{spverbatim}
\normalsize

We validated our prompt by manually evaluating the correcteness and the equivalence of 100 random questions and their 10 generated variations. The analysis show an accuracy of 100\% of the generative model. 
We reported some examples of the generated queries for both MSMARCO and NQ in Table~\ref{tab:query-examples}.

\begin{table}[t]
    \centering
    \begin{tabular}{p{0.95\linewidth}}
    \hline
    \textbf{MS-MARCO}\\
    \hline
    Q: \emph{What is the typical function of simple epithelium}\\
    G1: Could you explain the main function of simple epithelium?\\
    G2: What role does simple epithelium play in the body?\\
    G3: simple epithelium purpose\\
    \hline
    Q: \emph{what is federal prevailing wage}\\
    G1: Can you explain the concept of federal prevailing wage?\\
    G2: federal prevailing wage definition \\
    G3: What does federal prevailing wage refer to? \\
    G4: Can you explain what "federal prevailing wage" is? \\
    \hline
    \textbf{Natural Questions}\\
    \hline
    Q: \emph{when did the san francisco giants win their first world series} \\
    G1: When did the San Francisco Giants first win the World Series? \\
    G2: In what year did the San Francisco Giants first win the World Series? \\
    G3: ancisco Giants first win the World Series? When? \\
    \hline
    Q: \emph{when is shameless us season 8 coming out}\\
    G1: When is the eighth season of Shameless US scheduled to air? \\
    G2: If I'm looking for the premiere date of Shameless US Season 8, when should I expect it? \\
    G3: When will Shameless US Season 8 be available for viewing? \\
    \hline
    \end{tabular}
    \caption{Examples of original queries and their generated variations using Phi-3.}
    \label{tab:query-examples}
\end{table}

\section{STS training - ModernBERT}
\label{apx:sts}
Building upon from a public checkpoint of ModernBERT, 140M parameters, we performed extensive additional pre-training using diverse datasets focused on semantic text matching tasks. The training data encompassed various tasks including text similarity detection, answer matching, document understanding, and content abstraction. The training leveraged multiple large-scale datasets such as The Semantic Scholar Open Research Corpus~\cite{lo2020s2orc}, PAQ~\cite{lewis2021paq},  AmazonQA~\cite{https://doi.org/10.48550/arxiv.1908.04364},  WikiHow~\cite{https://doi.org/10.48550/arxiv.1810.09305}, and others. 
Overall, these resources contain more than $\approx1.5$B semantically related text pairs. 

Our training approach focused on semantic similarity learning, following established practices in dense retrieval training~\cite{reimers2019sentence}. The model was trained to distinguish between semantically related and unrelated text pairs. We implemented FP16 precision training with a MultipleNegativesRanking loss function, employing a learning rate of 2e-5. The training process utilized a batch size of 2048 samples, with input sequences capped at 128 tokens. We used 8x Nvidia H100 GPUs.

\begin{table*}
  \centering
  \begin{tabular}{lcccccc}
    \hline
    \multirow{2}{*}{\textbf{Configuration}} & \multicolumn{3}{c}{\textbf{ModernBERT-base}} & \multicolumn{3}{c}{\textbf{Pre-trained ModernBERT-base}}\\
     & \textbf{P@1} & \textbf{NDCG@10} & \textbf{RBO@5} & \textbf{P@1} & \textbf{NDCG@10} & \textbf{RBO@5}\\
    \hline
    \multicolumn{7}{c}{MS-MARCO} \\
    \hline
FT          &	    20.2    &37.5   &\ppm{0.33}{0.23} & 22.8 & 41.6 &\ppm{0.39}{0.25}\\

+ $\mathcal{L}_{CR}$ 
                      &     20.5    &37.83  &\ppm{0.52}{0.25} & \tr{23.0} & \tr{41.9} &\ppm{\tr{0.56}}{0.26}\\
    \hline
    \multicolumn{7}{c}{Natural Questions} \\
    \hline
    FT            &	25.7  &36.9   &\ppm{0.08}{0.13} & 36.6 & 50.4 &\ppm{0.15}{0.19}\\
+ $\mathcal{L}_{CR}$ 
                            &	30.4  &42.6   &\ppm{0.48}{0.24} & \tr{37.2} & \tr{51.1} &\ppm{\tr{0.65}}{0.22}\\
    \hline
  \end{tabular}
  \caption{Comparison of document relevance and coherence of ModernBERT and STS Pre-trained ModernBERT. RBO measures the coherence.}
  \label{tab:results_modern_bert_pt_vs_no_pt}
\end{table*}
\section{Effect of STS pre-training}
\label{apx:mbert}
In this section, we investigate the impact of semantic similarity pre-training on ModernBERT's performance. 
This comparison allows to understand how semantic similarity knowledge acquired during STS pre-training affects both relevance ranking and coherence of the rank.

Results in Table~\ref{tab:results_modern_bert_pt_vs_no_pt} demonstrate clear advantages of using pre-trained ModernBERT-base compared to its non-pre-trained counterpart across both MS-MARCO and NQ datasets. On MS-MARCO, the pre-trained model with standard fine-tuning (FT) achieved notable improvements in all metrics, with P@1 increasing from 20.2\% to 22.8\%, NDCG@10 from 37.5 to 41.6, and RBO@5 from 0.33 to 0.39. When applying our proposed $\mathcal{L}_{CR}$ loss, the pre-trained model maintained its superior performance, showing further marginal improvements across all metrics (P@1: 23.0\%, NDCG@10: 41.9, RBO@5: 0.56).

The advantages of pre-training are even more pronounced on the Natural Questions dataset, where the pre-trained model demonstrated substantial gains in effectiveness. With standard fine-tuning, pre-training improved P@1 by approximately 11 percentage points (from 25.7\% to 36.6\%) and NDCG@10 by 13.5 points (from 36.9 to 50.4). The addition of $\mathcal{L}_{CR}$ loss further enhanced these results, with the pre-trained model achieving the best overall performance (P@1: 37.2\%, NDCG@10: 51.1, RBO@5: 0.65). Notably, RBO@5 showed substantial improvement with pre-training, particularly when combined with $\mathcal{L}_{CR}$, suggesting that pre-training helps the model develop more consistent and coherent ranking behavior.

\section{BEIR and TREC-DL evaluation}
\label{apx:beir}
To assess the generalization capabilities of our model and its performance across diverse domains, we conducted extensive experiments using TREC-DL '19\footnote{\url{https://microsoft.github.io/msmarco/TREC-Deep-Learning-2019}} and '20\footnote{\url{https://microsoft.github.io/msmarco/TREC-Deep-Learning-2020.html}} benchmarks, and 
11 datasets from the BEIR benchmark. The results, presented in Table \ref{tab:results_trec} and \ref{tab:beir}, demonstrate interesting patterns across different configurations. 

On TREC benchmarks, our proposed approach achieves best NDCG on both versions of TREC test data, +0.68 and +0.21 compared to the strongest baseline, +1.37 and + 0.30 compared to the simple FT model. 

Our $\mathcal{L}_{CR}$ configuration achieved the best overall performance with an average score of 44.94 across all BEIR datasets, showing consistent improvements over the public checkpoint (43.56) and standard fine-tuning (44.46). Notable improvements were observed on several key datasets as Scifact, HotPotQA, Arguana, and NFCorpus, demonstrating enhanced capability in handling complex, multi-hop questions as well as maintaining robust retrieval capabilities across specialized content.

Interestingly, while the Full configuration, showed strong performance on specific datasets like HotpotQA (55.22) and Quora (88.87), it didn't achieve the best overall average (42.96). This suggests that the combination of all components might lead to some interference effects in certain domains.
We used a model trained on MS-MARCO. Thus, we only focus on the average result metric as individual benchmarks would require fine-tuning.

\begin{table}
  \centering
  \begin{tabular}{lcc}
    \hline
    \textbf{Configuration}  & \textbf{NDCG@10} & \textbf{RBO@5} \\
    \hline
    \multicolumn{3}{c}{TREC-DL '19} \\
    \hline
    Public ckpt &  64.35 & \ppm{14.07}{0.12}\\
    FT          &	 69.77 & \ppm{14.57}{0.13}\\
    + Gen. Qs   &   70.46 & \ppm{15.59}{0.15}\\
    + $\mathcal{L}_{QQ}$ & 69.39 & \ppm{16.07}{0.14}\\
    + $\mathcal{L}_{CR}$ & 71.14 & \ppm{19.47}{0.14}\\
    \hline
    \multicolumn{3}{c}{TREC-DL '20} \\
    \hline
    Public ckpt & 63.36 & \ppm{15.56}{0.13}\\
    FT          & 65.52 & \ppm{16.11}{0.14}\\
    + Gen. Qs   & 65.49 & \ppm{16.35}{0.15}\\
    + $\mathcal{L}_{QQ}$  & 65.61 & \ppm{16.30}{0.14}\\
    + $\mathcal{L}_{CR}$  & 65.82 & \ppm{19.09}{0.12} \\
    \hline

  \end{tabular}
  \caption{Results on TREC DL benchmarks.}
  \label{tab:results_trec}
\end{table}

\begin{sidewaystable*}
\begin{tabular}{lcccccccccccc}
\hline
\textbf{Configuration} & \rotatebox{270}{\textbf{Scifact}} & \rotatebox{270}{\textbf{Arguana}} & \rotatebox{270}{\textbf{Nfcorpus}} & \rotatebox{270}{\textbf{Webis-touche2020}} & \rotatebox{270}{\textbf{Fiqa}} & \rotatebox{270}{\textbf{Dbpedia-entity}} & \rotatebox{270}{\textbf{Climate-Fever}} & \rotatebox{270}{\textbf{Quora}} & \rotatebox{270}{\textbf{Scidocs}} & \rotatebox{270}{\textbf{Trec-covid}} & \rotatebox{270}{\textbf{HotPotQA}} & \rotatebox{270}{\textbf{AVG}} \\   
\hline
Public ckpt                             & 59.98                                & 51.01                                & 32.11                                 & 16.55                                         & 47.42                             & 35.31                                       & 25.35                                      & 87.69                              & 17.39                                & 58.77                                   & 47.57                                 & 43.56                            \\
FT                               & 59.08                                & 47.72                                & 31.94                                 & 23.89                                         & 46.89                             & 36.72                                       & 25.16                                      & 88.15                              & 16.88                                & 60.60                                   & 52.04                                 & 44.46                            \\
+ Gen. Qs                           & 59.56                                & 46.86                                & 32.26                                 & 23.82                                         & 44.42                             & 35.86                                       & 26.51                                      & 86.81                              & 16.65                                & 58.48                                   & 53.54                                 & 44.07                            \\
+ $\mathcal{L}_{QQ}$                & 59.56                                & 46.74                                & 31.93                                 & 24.52                                         & 46.38                             & 36.71                                       & 25.39                                      & 88.91                              & 16.88                                & 55.23                                   & 54.50                                 & 44.25                            \\
+ $\mathcal{L}_{CR}$                           & 60.46                                & 48.50                                & 32.77                                 & 23.54                                         & 46.16                             & 36.62                                       & 26.04                                      & 88.61                              & 17.24                                & 60.26                                   & 54.19                                 & \tr{44.94}                            \\
Full                       & 59.95                                & 45.42                                & 32.61                                 & 23.88                                         & 43.18                             & 36.53                                       & 25.95                                      & 88.87                              & 16.99                                & 44.00                                   & 55.22                                 & 42.96          \\       
\hline
\end{tabular}
  \caption{Results on BEIR benchmark. The model, MPNet, is trained on MS-MARCO. Dataset as CQAdupstack, BioASQ, Signal1m, Trec-news, Robust04 are not included sice they were not available.}
  \label{tab:beir}
\end{sidewaystable*}

\section{Reformulation experiments}
\label{apx:reformulation}
As discussed in this paper, query reformualation is out of the scope of this work as it introduces strong drawbacks, including cost and latency, to the retrieval pipeline.
However, for completeness we ran two simple train-free reformulation approaches as additional baselines. 

The first approach, here indicated as \emph{Centroid}, was introduced by~\citet{kostric2024surprisingly}.
In short, the DR model first computes the embeddings of the original query $\textbf{e}_q$ and all of its $k$ reformulations $\textbf{e}_{r_1}\dots\textbf{e}_{r_k}$. 
Then, the embedding used to compute the similarity against indexed documents is defined as the average of all query embeddings $\frac{1}{k+1}(\textbf{e}_q+\sum_i \textbf{e}_{r_i})$. 
The motivation of the approach is that the centroid of these rewrites adds robustness to the DR model as the center of mass of multiple reformualtions will likely correspond better to the user's information need than a single rewrite. 
Note that, in the original work, a waighted average is used, where each reformulation has a score depending on the conversation history, not available on our task. 
As second baseline, we ran the DR model on all available reformulation and selected the documents with highest scores with respect to the reformulated query. By doing so, the model can select documents that receive low score with the original query but high score with a reformulation.

We tested these approaches on TREC-DL '19 and '20 benchmarks. Results are showed in Table~\ref{tab:results_reformulation}.

\begin{table}[h]
  \centering
  \begin{tabular}{lcc}
    \hline
    \textbf{Configuration}  & \textbf{P@1} &  \textbf{NDCG@10} \\
    \hline
    \multicolumn{3}{c}{TREC-DL '19} \\
    \hline
    No reformulation & 83.72 & 69.77 \\
    Centroid & 76.74 & 67.16 \\
    Best & 82.80 & 65.02 \\
    \hline
    \multicolumn{3}{c}{TREC-DL '20} \\
    \hline
    No reformulation & 81.48 & 65.52 \\
    Centroid & 78.22 & 61.68 \\
    Best & 80.95 & 65.47 \\
    \hline

  \end{tabular}
  \caption{Results of reformulation approaches on TREC DL benchmarks.}
  \label{tab:results_reformulation}
\end{table}

Both reformulation approaches show poor performance. We conjecture these methods highlight their benefits on conversational settings rather than single turn Q\&A. 
Other authors~\cite{wang2020deep} applied more complex techniques, where a reformulator model is trained and rewarded to generate queries to achieve higher retrieval performance. 
However, results showed a very limited improvement on TREC-DL benchmarks, less than 0.1\% NDCG@10.

\section{Retrieve and Generate}
\label{apx:randg}
In order to further explore the contribution of our DR models on downstream applications, we simulated a RAG pipeline. 

We used an LLM, Mistral-7B-Instruct-v0.2, to generate an answer while using the top-5 documents retrieved by various DR models. Specifically, we used the MPNet fine-tuned on MS-MARCO with MNRL and our CR losses. 
The LLM, evaluated on KILT benchmarks~\cite{petroni2020kilt}, showed an average improvement in accuracy by +0.4\%, from 60.0 (MNRL) to 60.4 (our CR loss).

This quick experiment is not mean to be exhaustive as the focus of this work is improving quality and coherence of DR models. This experiment is meant to provide an intuition of downstream effects in terms of accuracy. These results need to be explored further.

\end{document}